\ifdefined\XeTeXversion\else\ifdefined\directlua\else\pdfoutput=1\fi\fi % arXiv: pdfLaTeX (keep in the first 5 lines)
\documentclass[11pt]{article}
\usepackage{clain}
\usepackage{times}
\usepackage{latexsym}
\usepackage[T1]{fontenc}
\usepackage[utf8]{inputenc}
\usepackage{microtype}
\usepackage{inconsolata}
\usepackage{graphicx}
\usepackage{amsmath}
\usepackage{enumitem}
\usepackage{CJKutf8}

\title{Words Caught in the Heart: Simulating Person-Specific Expression for Self-Opaque Clients in Psychotherapy}
\date{October 23, 2026}
\author{Mengxi~Xiao\textsuperscript{1,2}, 
Jingjing~Chen\textsuperscript{3}, 
Ziyan~Kuang\textsuperscript{1,2}, 
Jiarong~Song\textsuperscript{1,2}, 
Wenhao~Zhao\textsuperscript{4}, 
Zuchen~Han\textsuperscript{1,2},
Yihuai~Xu\textsuperscript{1,2}, 
Xiyang~Huang\textsuperscript{1,2},
Jie~Gong\textsuperscript{1,2}, 
Yuechen~Jiang\textsuperscript{5},
Xueqing~Peng\textsuperscript{6},
Yankai~Chen\textsuperscript{7},
Xiaoyan~Qu\textsuperscript{3,\dag}, 
Min~Peng\textsuperscript{1,2,\dag},
Qianqian~Xie\textsuperscript{1,2,\dag}
}
\affiliation[1]{School of Artificial Intelligence, Wuhan University}
\affiliation[2]{Cognitive Language and Information Lab}
\affiliation[3]{Wuhan University}
\affiliation[4]{Huazhong Agricultural University}
\affiliation[5]{University of Manchester}
\affiliation[6]{Yale University}
\affiliation[7]{Mohamed bin Zayed University of Artificial Intelligence}
\authornote{\textsuperscript{\dag}Corresponding authors}

\correspondence{Qianqian Xie <xieq@whu.edu.cn>, Min Peng <pengm.whu.edu.cn>, Xiaoyan Qu <qxy625ch@whu.edu.cn>}
\metadata[Code]{\ghlink{whunextgen/Pandora}}
\metadata[Website]{\href{https://clain.org/Pandora/}{clain.org/Pandora/}}

\begin{document}

% The abstract goes before \maketitle so it is placed inside the title card.
\begin{abstract}
Psychotherapy clients may experience distress before they can coherently describe it, and their expression changes with emotion, perceived safety, and trust. Large language model-based client simulators often rely on clinically organized case descriptions, which can encourage responses explaining the client's difficulties more clearly than their current understanding warrants. We propose \textbf{Pandora}, a psychotherapy client simulation framework that models how psychological experience becomes situated, individual expression. Pandora updates psychological and relational states in response to therapist interventions, constructs subjective appraisals and attributions, and plans response content and strategy. A separate expression planner combines these decisions with language patterns distilled from the client's utterances to guide linguistic realization. We evaluate Pandora against five baselines on 115 cases from four counseling and motivational interviewing datasets. It achieves the highest expression similarity and overall disclosure-alignment score among the compared systems. In a four-evaluator study of 45 matched disclosure items, Pandora receives AI-like judgments in 7.2\% of ratings, compared with 43.3--67.2\% for the baselines and 35.6\% without expression planning. The ablation also reduces expression similarity. These results support explicit expression planning in client simulation to improve alignment with individual language patterns and reduce perceived AI-like expression in the evaluated setting.
\end{abstract}

\maketitle\par\color{black}

\section{Introduction}

\begin{figure*}[t]
\centering
\includegraphics[width=\textwidth]{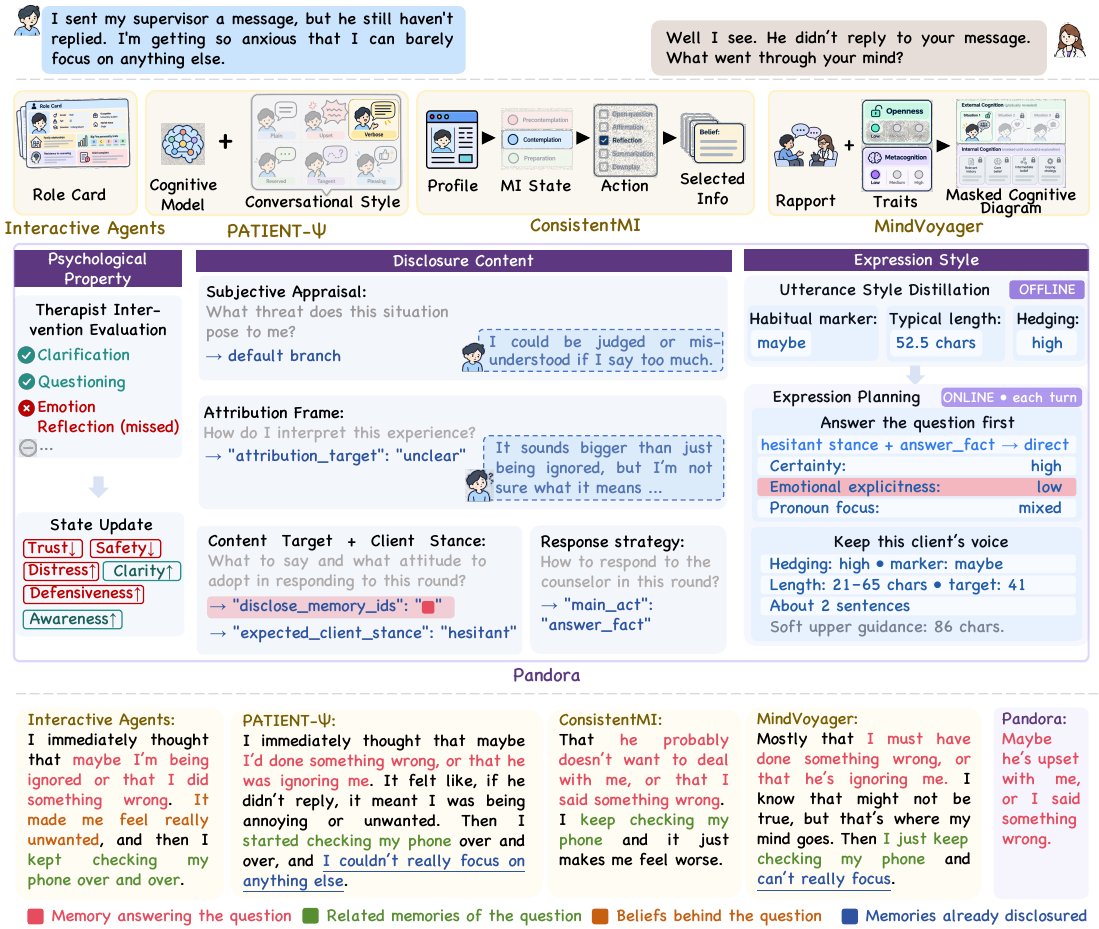}
\caption{An example of representative psychotherapy client simulators and our proposed framework, Pandora, responding to the same therapist utterance. In this case, the therapist misses emotional reflection, lowering the client's perceived safety and increasing defensiveness. Only Pandora explicitly captures this omission and its psychological consequences, producing a brief, tentative response.}
\label{fig:pipeline_comparison}
\end{figure*}

Realistic user simulation depends on capturing how people turn their experiences and intentions into language \citep{levelt1999lexical}. This requirement is particularly important in psychotherapy, where what a client says can shape the therapist’s understanding of the problem and the direction of the conversation \citep{stiles1990assimilation,pos2003processing,blanchard2016lying,safran1996ruptures}. Clients may seek help before they can give a coherent account of their difficulties, and their ability to express themselves can vary with emotional distress, perceived safety, and trust in the therapist \citep{beukeboom2006mood,farber2004disclosure,blanchard2016lying}. A concern may initially emerge through a concrete incident or a tentative remark, then become more explicit as the conversation develops \citep{gendlin1961experiencing,stiles1990assimilation,pos2003processing}. For simulators used in therapist training and evaluation, these differences matter: an unusually articulate client may make a difficult concern easier to identify and explore than it would be in practice \citep{dasilva2018alliance,kahn2008disclosures}. Credible simulation therefore requires attention to the relationship between a client’s psychological experience and its expression at a particular moment.

Existing client simulators model cognitive case formulations and conversational styles (Patient-$\Psi$), metacognition and openness (MindVoyager), and relational patterns and emotional reactions (PatientAct) \citep{wang2024patient,kim2025mindvoyager,sabour2026patientact}.  These representations provide useful structure, but their organization often reflects the clinician’s explanatory account of the client. A case formulation connects experiences, beliefs, and coping patterns into an interpretation that the client may only partially recognize or articulate \citep{eells1998formulation,persons2008formulation}. Using this formulation to guide generation can consequently encourage responses with greater explanatory coherence than the client’s current understanding warrants \citep{kim2025mindvoyager}. For example, when a client begins to recognize a recurring tendency to devalue personal achievements, that awareness need not immediately become a coherent explanation \citep{gendlin1961experiencing}. It may instead appear through a specific experience: “Whenever someone praises me, I feel uncomfortable. It doesn’t really feel like something I accomplished.” The underlying pattern is present, but its expression remains tied to the immediate situation and incomplete. A conversational style label may alter the tone of such a response while leaving its psychological explanation unusually explicit \citep{wang2024patient}.

% \footnote{Code for dialogue preprocessing, client and therapist distillation, and Pandora modeling: \url{https://anonymous.4open.science/r/Pandora-Modeling/}.}
To bridge this gap, we propose \textbf{Pandora}, a psychotherapy client simulation framework that models how psychological experience becomes individual expression. Drawing on the distinction between conceptual preparation and linguistic formulation in speech production \citep{levelt1999lexical}, Pandora explicitly plans both what a client communicates and how that client says it. As illustrated in Figure~\ref{fig:pipeline_comparison}, the framework first interprets the therapist’s intervention and updates the client’s psychological and relational states. It then constructs a subjective appraisal and attribution frame, capturing how the client experiences and understands the situation, and plans the response’s content, stance, and conversational purpose. A separate expression planner combines these decisions with language patterns distilled from the client’s own utterances to guide wording, hedging, emotional explicitness, and response length. In the illustrated exchange, an unanswered message becomes a brief conjecture, “Maybe he’s upset with me, or I said something wrong”, that conveys the client’s concern while preserving uncertainty about its meaning. This design connects evolving psychological states with a recognizable individual voice, making the transformation from subjective experience to situated speech an explicit part of client simulation.

We evaluate Pandora on 115 cases drawn from four counseling and motivational interviewing datasets, comparing it with five baselines and an ablation without expression planning. Automatic evaluation measures both disclosure alignment and expression fidelity against the corresponding source dialogues. A complementary human study assesses perceived AI-like expression on 45 matched disclosure items, with judgments from two counselors and two synthetic-data researchers. Pandora achieves the highest expression similarity and overall disclosure-alignment score among the compared systems. In human evaluation, its AI-like judgment rate is 7.2\%, compared with 43.3--67.2\% for the baselines. Removing expression planning raises this rate to 35.6\% and reduces expression similarity, highlighting the contribution of linguistic realization beyond content planning. Together, these findings support explicit expression planning as a central component of client simulation, improving fidelity to individual language patterns while reducing perceived AI-like expression.

Our contributions are summarized as follows:
\begin{itemize}[leftmargin=0em, itemindent=1em]
\item We identify the gap between clinically organized client representations and clients’ own expression, and formulate psychotherapy client simulation around how psychological experience becomes situated speech (Section~3).
\item We propose \textbf{Pandora}, which connects psychological state updates with subjective interpretation, content planning, and a separate linguistic realization plan. The framework combines the demands of the current interaction with habitual language patterns to produce responses that reflect both the client’s psychological state and individual voice (Section~3).
\item We develop an evaluation combining disclosure alignment, expression fidelity, and human judgments of AI-like expression. Comparisons with five baselines on 115 cases, complemented by human assessment of 45 matched disclosure items, show stronger expression fidelity and less perceived AI-like expression; the ablation further supports the contribution of expression planning (Section~4).
\end{itemize}

\section{Related Work}

\paragraph{Language Expression in User Simulation.}
Existing user simulators model expression through linguistic controls, user profiles, and internal states. Personality-oriented generation controls hedging, verbosity, and lexical and syntactic choices \citep{mairesse2011controlling}; mTAD models utterance traits such as emotion, fluency, and repetition, while USP infers personality and communication styles from conversations \citep{ferreira2024multitrait,wang2025implicitprofiles}. More recent approaches reconstruct unexpressed thoughts or align cognitive, affective, and communication states before generating responses \citep{wu2026thinking,wu2026humanlm}. Psychotherapy adds a specific modeling demand: how a client's evolving understanding and experience of the interaction shape the expression of a concern. A concern may be felt before it can be coherently explained \citep{gendlin1961experiencing}, and willingness to disclose it can be shaped by perceived safety and trust \citep{farber2004disclosure}. Style controls and internal-state descriptions do not by themselves specify how these conditions jointly shape momentary expression. Pandora takes the client's subjective interpretation as the starting point for expression planning. Subjective appraisal and attribution guide content planning, while a separate linguistic realization plan expresses that perspective through the client's habitual language patterns under the current psychological and interactional conditions.

\paragraph{LLM-based Psychotherapy Client Simulation.}
Existing client simulators seek realism through psychological profiles and mechanisms that regulate counseling behavior. Patient-$\Psi$ combines CBT case formulations with conversational styles, while Interactive Agents grounds simulated clients in expert-written role cards \citep{wang2024patient,qiu2024interactive}. MindVoyager models cognitive recognition and disclosure through metacognition, openness, and rapport, whereas ConsistentMI separates readiness for behavioral change, action selection, and response generation \citep{kim2025mindvoyager,yang2025consistent}. AnnaAgent models evolving emotions and complaints, and Eeyore learns depression-related client expression through expert-guided supervised and preference optimization \citep{wang2025annaagent,liu2025eeyore}. PatientAct connects clinical and relational formulations to emotional reactions and behaviors, expressing resistance through quantity, content, and style; PsyCLIENT guides responses through behavioral trajectories and reference content \citep{sabour2026patientact,qiu2026psyclient}. These methods provide clinically grounded content and mechanisms for shaping interactional behavior. Our focus is how this material becomes the client's own account: an explanatory pattern available to the simulator may remain only partly understood by the client. Pandora constructs a situated subjective interpretation through appraisal and attribution, then plans its content and linguistic realization using the client's habitual language patterns. This explicitly targets the relationship between psychological understanding and individual expression in client simulation.

% \paragraph{Position of Our Work.}
% Pandora brings these components together around the realization of psychological experience as situated client speech. Its focus is the transformation from the simulator's explanatory representation to the client's current subjective interpretation, and from that interpretation to a conversational response. Subjective appraisal and attribution inform content and response-act planning; a distinct linguistic realization plan then combines these decisions with psychological state and individual utterance patterns to specify certainty, hedging, emotional explicitness, and response length. This decomposition makes the relationship between what a client understands, what the current exchange calls for, and how that client expresses it an explicit part of generation.

\section{Method}
\label{sec:method}

Pandora organizes client simulation around the transformation of psychological experience into situated speech. Following the distinction between conceptual preparation and linguistic formulation in speech production \citep{levelt1999lexical}, the framework connects three modeling tasks: representing the client's evolving experience of the interaction, planning a response from the client's subjective perspective, and realizing that response in the client's individual voice. Figure~\ref{fig:pipeline_comparison} shows the modular process: intervention evaluation and state update provide the context for subjective appraisal and attribution; content and response strategy planning organize the reply; and individual expression planning guides its linguistic realization.

At turn $t$, the inputs are the therapist utterance $x_t$, preceding dialogue $H_{t-1}$, previous psychological state $s_{t-1}$, case material $M_t$, and an individual language profile $\rho$. The case material $M_t$ contains memories and events available at the current dialogue point, with future events excluded according to their occurrence time. All psychological depths in this material are available to planning. This separates the case information available to the simulator from the client's evolving understanding and manner of expressing it.

\subsection{Therapist Intervention Evaluation and State Update}

This module provides a persistent representation of the client's experience of the interaction \citep{hill1993covert,safran1996ruptures}. A rule-based evaluator identifies opportunities for clarification, questioning, and emotional reflection, together with their use in the therapist's utterance. The resulting intervention signal represents both the techniques used and how they address the conversational context, including opportunities left unaddressed.

The state distinguishes affective, relational, defensive, and reflective dimensions, allowing willingness to disclose and capacity to articulate an experience to vary separately \citep{farber2004disclosure,grant2002reflection}. The full representation is specified in Appendix~\ref{app:implementation}. Bounded rule-based transitions combine intervention-related and memory-related increments with the preceding state. These dimensions jointly guide subsequent planning: a client can gain clarity about a concern while remaining hesitant to discuss it. Figure~\ref{fig:pipeline_comparison} illustrates this combination of cognitive and relational responses. Disclosure readiness shapes the client's stance, while the supplied case material remains available. Update rules and numerical settings are provided in Appendix~\ref{app:implementation}.

\subsection{Subjective Appraisal and Attribution}

This module turns the updated state and case material into the perspective from which the client speaks. Subjective appraisal $a_t$ represents what an experience means to the client, including its emotional significance, perceived threat, implications for the self and relationships, and confidence in that interpretation. Attribution $b_t$ represents how the client currently explains the experience. Both are constructed using the updated state, case material, and intervention signal.

Appraisal and attribution distinguish the explanatory information in a case from the account the client can currently formulate. State-conditioned attribution ranges from immediate behavior and circumstances to tentative relational interpretations and connections with underlying beliefs. The intervention signal further conditions how clearly this account is formulated. Thus, the same case material can support different subjective accounts as the interaction develops, preserving uncertainty even when the client begins to understand a concern more clearly.

\subsection{Content Target, Client Stance, and Response Strategy}

This module organizes the subjective account into a response to the current therapist turn. A content plan $d_t$ represents available case content and client stance, while a response-act plan $r_t$ specifies the immediate conversational purpose:
\begin{equation}
C_t=(a_t,b_t,d_t,r_t).
\end{equation}
The content plan carries the supplied memory identifiers together with a stance such as hesitant, defensive, confused, emotionally overwhelmed, or open. The updated state and intervention signal determine this stance, linking case content to the client's current engagement with the conversation.

The response strategy makes the plan responsive to the therapist's utterance. Factual questions favor an answer first with limited introspection; practical requests favor brevity; reflections invite confirmation or refinement; and challenges or invalidation allow resistance or clarification. Separating response purpose from stance allows a hesitant client to answer a question directly while expressing uncertainty about its psychological meaning. These decisions form the semantic input to expression planning.

\subsection{Individual Expression Planning}

Expression planning translates the semantic plan into realization guidance for the particular client. It combines a stable individual language profile distilled offline from the client's utterances with the psychological and conversational conditions of the current turn:
\begin{equation}
L_t=F_{\mathrm{realize}}(s_t,C_t,H_{t-1},\rho).
\end{equation}
This interface makes individual voice and the current interaction explicit inputs to realization.

\paragraph{Offline utterance style distillation.}
The language profile summarizes habitual discourse markers, hedging, emotional explicitness, utterance length distributions, and sentence counts. Length statistics describe central tendencies and distributional ranges, while marker statistics capture expressions recurring in the client's own language. Reusing this profile across turns provides concrete guidance for maintaining individual voice.

\paragraph{Online expression planning.}
The planner specifies speech style, pronoun focus, certainty, emotional explicitness, hedging, and length and sentence-count targets. These variables draw on research on psychologically informative word use \citep{pennebaker2003language}, controllable stylistic generation \citep{mairesse2011controlling}, hedging under uncertainty \citep{smith1993answering}, and the communicative functions of filled pauses \citep{clark2002fillers}. Client stance establishes a base speech style, attribution guides focus on the self, others, or the situation, and response strategy adapts these choices to the current exchange. An answer-first strategy, for example, favors direct wording and low emotional explicitness, while resistance or uncertain interpretation favors tentative expression. The individual profile separately conditions habitual hedging, allowing a direct answer to retain a familiar marker such as ``maybe.'' This combination preserves the client's voice while adapting expression to the current exchange.

\subsection{Response Generation and Execution}

The client LLM realizes the intermediate plans as a natural-language utterance. Its prompt combines the psychological context, subjective appraisal and attribution, content and response strategy plans, and linguistic realization plan with the client profile, interaction instructions, current therapist utterance, and dialogue history. The model is instructed to answer the current request from the client's perspective and follow the individual realization guidance, with internal identifiers and numerical state fields kept out of the utterance. Explicit rules organize the intermediate plans, and the LLM supplies their linguistic realization.

Content targeting and claim verification track the expression of case material across turns, as specified in Appendix~\ref{app:implementation}. The resulting dialogue context supports the next planning cycle. A stage-by-stage running example is provided in Appendix~\ref{app:example}.

\section{Experiment}

We evaluate Pandora's central design choice: explicitly planning how a client's subjective perspective takes linguistic form in the current exchange. The experiments address three questions: \textbf{RQ1}, does Pandora better reproduce individual expression patterns and reduce perceived AI-like expression? \textbf{RQ2}, what does the expression planner contribute beyond the preceding psychological and content planning stages? \textbf{RQ3}, are these expression gains accompanied by faithful disclosure timing and structure? We use automatic comparisons on 115 cases and human judgments of matched disclosures from 45 cases. The full-system comparisons address expression fidelity, the expression-planning ablation assesses the contribution of the realization stage, and disclosure metrics examine how case material unfolds in the generated conversation.

\subsection{Data Collection and Case Selection}
\label{sec:data_collection}

Evaluating client expression requires source material that reveals both what a client discloses and how that disclosure unfolds through interaction. The retrieval and assessment procedure is detailed in Appendix~\ref{app:case_selection}.

After semantic screening and removal of exact and near duplicates, the final collection contains 115 cases: 97 Chinese and 18 English, comprising 97 PsyDial-D4, 11 AnnoMI, five Eeyore Full, and two ConsistentMI cases. The selected dialogues provide reference trajectories for evaluating disclosure timing, content organization, and individual expression across turns.

Table~\ref{tab:case_sources} summarizes the source composition.

\begin{table}[t]
\centering
\small
\begin{tabular}{lr}
\hline
Dataset & Selected cases \\
\hline
\href{https://huggingface.co/datasets/qiuhuachuan/PsyDial-D4}{PsyDial-D4} & 97 \\
\href{https://github.com/uccollab/AnnoMI}{AnnoMI} & 11 \\
\href{https://github.com/MichiganNLP/Eeyore}{Eeyore Full} & 5 \\
\href{https://github.com/IzzetYoung/ConsistentMIClientSimulator}{ConsistentMI} & 2 \\
\hline
Total & 115 \\
\hline
\end{tabular}
\caption{Composition of the final case collection after semantic screening and cross-source deduplication. Dataset names link to their official releases.}
\label{tab:case_sources}
\end{table}

\subsection{Evaluation Framework}
\label{sec:evaluation_metrics}

We evaluate two complementary aspects of client simulation: disclosure content (RQ3) and expression style (RQ1 and RQ2). Content evaluation examines how case material unfolds across turns, while style evaluation combines textual similarity with direct human judgments of AI-like expression.

\subsubsection{Disclosure Content}
We use a shared memory-card codebook with levels L1--L3 and compare the first disclosure of each card in the reference and generated dialogues. For controller-trace inputs, disclosure is determined by evidence-grounded claim verification: a card is completed when all its required claims have been supported cumulatively by client utterances. A generator's self-reported disclosure depth is not used as the card-disclosure observation. LDDA measures disclosure timing, DS measures prerequisite satisfaction, BFE and BSE measure bundling errors, and GDS aggregates timing, dependency, and bundling fidelity. The complete definitions and conventions are provided in Appendix~\ref{app:metrics}.

\subsubsection{Expression Style}
\paragraph{Automatic evaluation.}
Following the stylometric approach used in Eval4Sim~\citep{bao2026eval4sim}, we represent client utterances using character 4-gram TF-IDF with up to 4,000 features and cosine similarity. We adapt this representation to paired comparisons between generated client speech and its corresponding source dialogue, rather than reproducing Eval4Sim's trained authorship verifier or aggregate PAN consistency score. To distinguish expression patterns from shared subject matter, we compute both raw and content-reduced similarities. Full weighting and matched-block definitions are provided in Appendix~\ref{app:metrics}. We report case-averaged raw and content-reduced similarities separately, with larger values indicating closer textual expression. The current 115-case evaluation fits a separate reference instrument for each case and aggregates matched similarities; it does not compute cross-client retrieval scores. The masking lexicon is Chinese-oriented and compresses Latin-script words into generic masks, so the content-reduced view provides coarser evidence for English cases. Neither view establishes psychological authenticity or clinical equivalence, and their absolute difference is not interpreted as a measured content effect.

\paragraph{Human evaluation.}
\label{sec:human_evaluation}
We sampled 45 disclosure items from distinct cases, with 15 items at each memory level (L1--L3). Each item presents eight candidates in randomized order: the reference, five baselines, Pandora, and its expression-planning ablation. Four evaluators (two counselors and two synthetic-data researchers) independently judged every candidate using a binary response: $y_{isr}=1$ for AI-like expression and $0$ otherwise, for item $i$, source $s$, and evaluator $r$. We report the AI-like judgment rate (AIR) as the mean of these binary judgments, yielding 180 judgments per source. Across all 360 candidates, mean pairwise agreement is 62.9\%, and positive agreement for AI-like judgments is 53.7\% (95\% CI: 49.4--58.1\%). Full presentation, annotation guidelines, aggregation, and uncertainty details are provided in Appendix~\ref{app:human_protocol}.

\subsection{Experimental Results}
\label{sec:experimental_results}

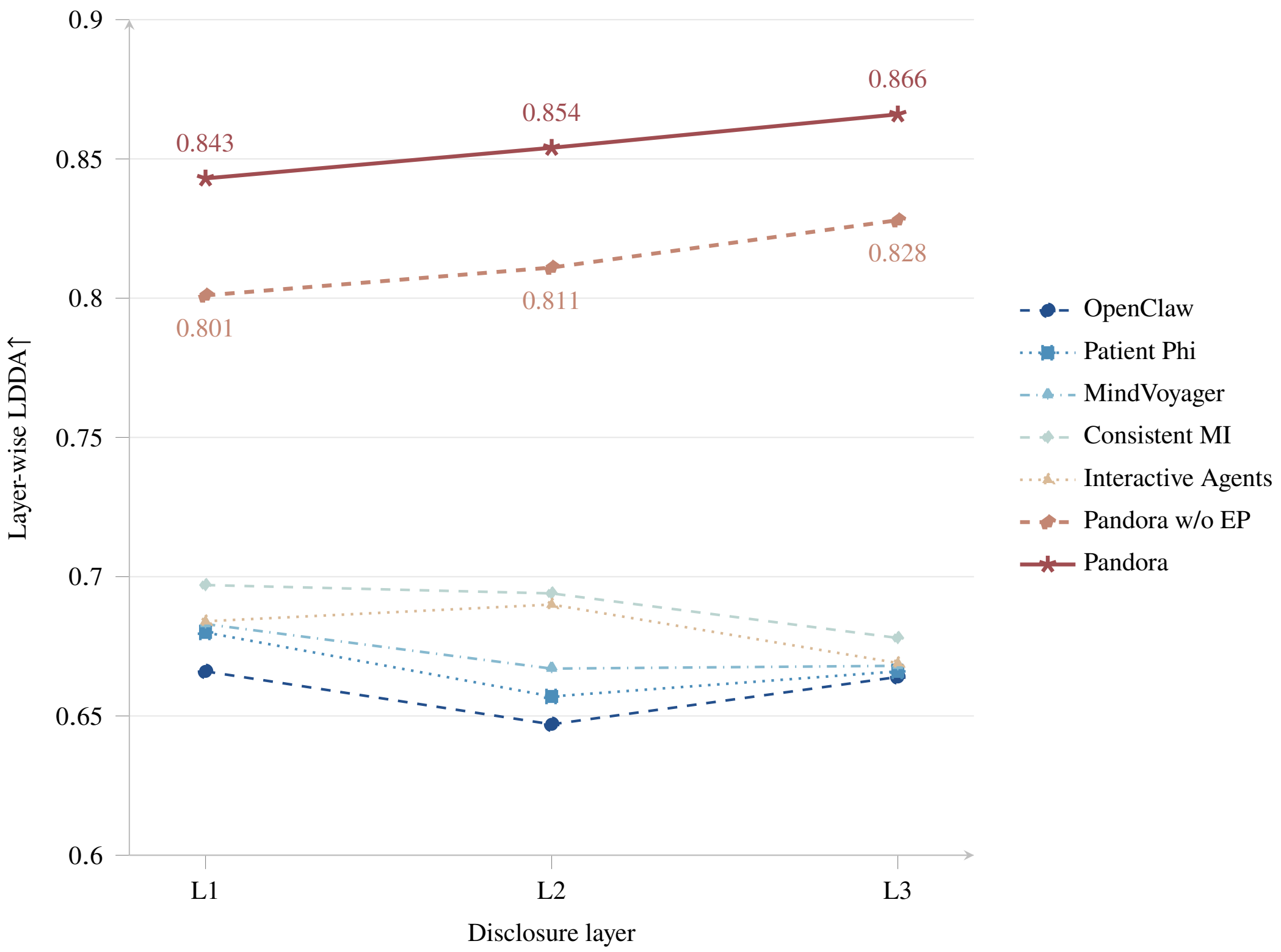
\begin{figure}[t]
\centering
\begin{tikzpicture}
\definecolor{ldda0}{HTML}{234f8c}
\definecolor{ldda1}{HTML}{4c8eba}
\definecolor{ldda2}{HTML}{86b9cf}
\definecolor{ldda3}{HTML}{bbd4d0}
\definecolor{ldda4}{HTML}{d9ba97}
\definecolor{ldda5}{HTML}{c38673}
\definecolor{ldda6}{HTML}{a04d51}
\begin{axis}[
width=0.65\columnwidth,height=0.644\columnwidth,
xmin=0.78,xmax=3.22,ymin=0.60,ymax=0.90,
xtick={1,2,3},xticklabels={L1,L2,L3},ytick={0.60,0.65,0.70,0.75,0.80,0.85,0.90},
ylabel={Layer-wise LDDA$\uparrow$},xlabel={Disclosure layer},
axis lines=left,tick align=outside,axis line style={gray!50},
ymajorgrids=true,grid style={gray!18},tick label style={font=\scriptsize},label style={font=\scriptsize},
legend style={at={(1.04,0.5)},anchor=west,draw=none,font=\scriptsize,legend columns=1,row sep=2pt},
legend cell align=left,clip=false]
\addplot[color=ldda0,dashed,line width=0.8pt,mark=*,mark size=2pt] coordinates {(1,0.666) (2,0.647) (3,0.664)};
\addlegendentry{OpenClaw}
\addplot[color=ldda1,dotted,line width=0.8pt,mark=square*,mark size=2pt] coordinates {(1,0.680) (2,0.657) (3,0.666)};
\addlegendentry{Patient Phi}
\addplot[color=ldda2,dashdotted,line width=0.8pt,mark=triangle*,mark size=2pt] coordinates {(1,0.683) (2,0.667) (3,0.668)};
\addlegendentry{MindVoyager}
\addplot[color=ldda3,dashed,line width=0.8pt,mark=diamond*,mark size=2pt] coordinates {(1,0.697) (2,0.694) (3,0.678)};
\addlegendentry{Consistent MI}
\addplot[color=ldda4,dotted,line width=0.8pt,mark=triangle*,mark size=2pt] coordinates {(1,0.684) (2,0.690) (3,0.669)};
\addlegendentry{Interactive Agents}
\addplot[color=ldda5,dashed,line width=1.2pt,mark=pentagon*,mark size=2pt] coordinates {(1,0.801) (2,0.811) (3,0.828)};
\addlegendentry{Pandora w/o EP}
\node[font=\scriptsize,text=ldda5,anchor=north] at (axis cs:1,0.797) {0.801};
\node[font=\scriptsize,text=ldda5,anchor=north] at (axis cs:2,0.807) {0.811};
\node[font=\scriptsize,text=ldda5,anchor=north] at (axis cs:3,0.824) {0.828};
\addplot[color=ldda6,solid,line width=1.2pt,mark=star,mark size=2.7pt] coordinates {(1,0.843) (2,0.854) (3,0.866)};
\addlegendentry{Pandora}
\node[font=\scriptsize,text=ldda6,anchor=south] at (axis cs:1,0.848) {0.843};
\node[font=\scriptsize,text=ldda6,anchor=south] at (axis cs:2,0.859) {0.854};
\node[font=\scriptsize,text=ldda6,anchor=south] at (axis cs:3,0.871) {0.866};
\end{axis}
\end{tikzpicture}
\caption{Layer-wise LDDA averaged over cases with reference-observed cards (L1: $n=113$; L2: $n=115$; L3: $n=88$). Pandora leads at every layer. Lines connect layer means; EP denotes expression planning.}
\label{fig:layer_ldda}
\end{figure}

We compare seven implementations on the same 115 cases. The evaluated baseline adaptations span direct role-play (OpenClaw), role-card context (Interactive Agents), cognitive-model context with style guidance (Patient Phi), cognitive and relational conditioning (MindVoyager), and state-conditioned action selection (Consistent MI). Pandora connects state updates to subjective appraisal, attribution, and response planning, followed by a distinct individual expression plan. Its ablation retains the preceding stages while removing the expression module. Table~\ref{tab:main_results} reports unweighted case means, and Figure~\ref{fig:layer_ldda} gives timing alignment by memory level. We examine which aspects of this organization contribute to disclosure content and expression style.

\begin{table*}[t]
\centering
\small
\setlength{\tabcolsep}{5pt}
\begin{tabular}{lr|rrr|r|rr}
\toprule
& \multicolumn{5}{c|}{Memory disclosure} & \multicolumn{2}{c}{Expression fidelity} \\
\cmidrule(lr){2-6}\cmidrule(lr){7-8}
Method & \textbf{LDDA} $\uparrow$ & DS $\uparrow$ & BFE $\downarrow$ & BSE $\downarrow$ & \textbf{GDS} $\uparrow$ & $S_{\mathrm{style}}$ $\uparrow$ & $S_{\mathrm{raw}}$ $\uparrow$ \\
\midrule
OpenClaw & 0.658 & 0.362 & 0.115 & 0.270 & 0.609 & 0.397 & 0.280 \\
Patient Phi & 0.667 & 0.373 & 0.108 & 0.200 & 0.629 & 0.438 & 0.310 \\
MindVoyager & 0.672 & 0.378 & 0.105 & 0.190 & 0.634 & 0.408 & 0.290 \\
Consistent MI & 0.690 & 0.417 & 0.100 & 0.128 & 0.664 & 0.512 & 0.352 \\
Interactive Agents & 0.681 & 0.364 & \textbf{0.099} & 0.133 & 0.643 & 0.501 & 0.342 \\
\midrule
Pandora w/o EP & 0.811 & 0.795 & 0.107 & 0.140 & 0.828 & 0.458 & 0.311 \\

Pandora & \textbf{0.852} & \textbf{0.875} & 0.102 & \textbf{0.127} & \textbf{0.871} & \textbf{0.566} & \textbf{0.385} \\
\bottomrule
\end{tabular}
\caption{Case-averaged results on Core115. $\mathrm{LDDA}_{\mathrm{macro}}$ summarizes disclosure timing across layers; DS and BFE/BSE measure dependency and bundling behavior. GDS combines timing, dependency, and bundling fidelity. Expression scores use content-reduced ($S_{\mathrm{style}}$) and raw ($S_{\mathrm{raw}}$) text. Bold marks the best unrounded mean in each column. EP: expression planning.}
\label{tab:main_results}
\end{table*}

\paragraph{Connecting psychological context to expression (RQ1).}
State-based baselines differ substantially in expression fidelity. Consistent MI, which conditions response actions on its state, is the strongest baseline on both similarity measures. MindVoyager's cognitive and relational conditioning yields a content-reduced similarity of 0.408, below the role-card-based Interactive Agents at 0.501. This ordering suggests that psychological context alone does not ensure fidelity to a client's manner of speaking. Pandora explicitly carries the client's subjective interpretation and response purpose into individual realization, achieving content-reduced/raw similarities of 0.566/0.385, exceeding Consistent MI by 0.053/0.032. The same ordering holds on the 97 Chinese cases: 0.489/0.422 for Pandora, 0.427/0.385 for Consistent MI, and 0.362/0.337 for the expression ablation. The advantage therefore persists where the content-reduction procedure is best matched to the language. The design implication is to specify how psychological context shapes wording, certainty, and utterance form through an explicit realization interface.

\paragraph{Content planning and expression planning serve distinct roles (RQ2).}
The ablation provides the clearest evidence for this separation. Without expression planning, Pandora retains a GDS of 0.828, above the strongest baseline's 0.664, but its content-reduced similarity falls to 0.458, below Consistent MI. Thus, the upstream state, subjective interpretation, and response plans can organize disclosure effectively while leaving individual expression insufficiently specified. Restoring the expression module raises content-reduced similarity from 0.458 to 0.566 and raw similarity from 0.311 to 0.385, gains of 0.108 and 0.074. LDDA, DS, and GDS also improve by 0.042, 0.079, and 0.043. These descriptive comparisons support a modular design in which semantic planning organizes the client's account and a separate realization stage preserves individual voice. The ablation evaluates the expression module as a whole, combining the individual language profile with its conditioning on the current interaction.

\paragraph{Coordinated planning supports disclosure organization (RQ3).}
The content-side results complement the expression gains. Pandora achieves a GDS of 0.871, compared with 0.664 for Consistent MI and 0.828 for the ablation, with timing alignment of 0.852 and dependency satisfaction of 0.875. Its timing advantage spans L1--L3 (0.843, 0.854, and 0.866), exceeding every baseline and the ablation at each level (Figure~\ref{fig:layer_ldda}). This pattern supports organizing available case material through a situated subjective account and response plan, alongside planning its expression. The strongest gains concern when material is disclosed and how supporting disclosures precede deeper ones. Bundling reflects a different tradeoff: Interactive Agents has the lowest frequency error (0.099 versus Pandora's 0.102), while Pandora has the lowest size error (0.127 versus Consistent MI's 0.128). Effective simulation therefore calls for coordinated planning of disclosure timing, dependencies, and individual expression; brevity or low bundling error alone captures only part of that goal.

\paragraph{Individual realization improves perceived client voice (RQ1 and RQ2).}
Human judgments show that the realization difference is perceptible (Table~\ref{tab:human_ai}). Among the baselines, Interactive Agents has the lowest AIR, whereas Consistent MI leads the automatic expression metrics: textual similarity and perceived AI-like expression capture complementary properties. Pandora improves both, receiving 13 AI-like votes out of 180 (7.2\%), compared with 43.3--67.2\% for the baselines, a 36.1-point reduction relative to Interactive Agents. Removing expression planning raises AIR to 35.6\%, a paired difference of 28.3 points (exploratory case-bootstrap 95\% CI: 20.6--36.1 points). Together with the similarity results, this contrast supports explicitly realizing the planned account through the client's habitual language patterns and current interactional stance. A well-organized psychological account benefits from a dedicated mechanism for expressing it in a recognizable client voice.

\begin{table}[t]
\centering\small
\begin{tabular}{lrr}
\toprule
Source & AI votes / 180 & AIR $\downarrow$ \\
\midrule
OpenClaw & 111 & 61.7\% \\
Patient Phi & 121 & 67.2\% \\
MindVoyager & 98 & 54.4\% \\
Consistent MI & 85 & 47.2\% \\
Interactive Agents & 78 & 43.3\% \\
Pandora w/o EP & 64 & 35.6\% \\
Pandora & 13 & \textbf{7.2\%} \\
\midrule
Reference & 8 & 4.4\% \\
\bottomrule
\end{tabular}
\caption{Binary AI-like judgment rates from four evaluators (two counselors and two synthetic-data researchers), with 45 candidates and 180 judgments per source. Bold marks the lowest rate among generated systems. Reference retains source provenance and is not a uniformly verified human control. EP: expression planning.}
\label{tab:human_ai}
\end{table}

% \begin{table}[t]
% \centering
% \small
% \setlength{\tabcolsep}{3pt}
% \begin{tabular}{@{}p{0.40\columnwidth}rrr|r@{}}
% \toprule
% Method & L1 $\uparrow$ & L2 $\uparrow$ & L3 $\uparrow$ & \textbf{LDDA} $\uparrow$ \\
% \midrule
% OpenClaw & 0.666 & 0.647 & 0.664 & 0.658 \\
% Patient Phi & 0.680 & 0.657 & 0.666 & 0.667 \\
% MindVoyager & 0.683 & 0.667 & 0.668 & 0.672 \\
% Consistent MI & 0.697 & 0.694 & 0.678 & 0.690 \\
% Interactive Agents & 0.684 & 0.690 & 0.669 & 0.681 \\ \midrule
% Pandora w/o EP & 0.801 & 0.811 & 0.828 & 0.811 \\
% Pandora & \textbf{0.843} & \textbf{0.854} & \textbf{0.866} & \textbf{0.852} \\
% \bottomrule
% \end{tabular}
% \caption{Layer-wise LDDA and its overall timing score. Each layer is averaged over cases with reference-observed cards at that layer: L1, $n=113$; L2, $n=115$; L3, $n=88$. $\mathrm{LDDA}_{\mathrm{macro}}$ first averages the nonempty layers within each case and then averages all 115 cases; it is not the arithmetic mean of the three displayed column means. EP denotes expression planning.}
% \label{tab:layer_ldda}
% \end{table}

Pandora also has the lowest generated-system AIR in both evaluator groups: 10.0\% among counselors and 4.4\% among synthetic-data researchers, compared with 40.0\% and 31.1\% for the ablation. With two evaluators per group, these are descriptive comparisons. No Pandora candidate receives a strict majority of AI-like votes (at least three of four): 35 of the 45 candidates receive none, seven receive one, and three receive two. The reduction therefore extends across the sampled disclosures as well as both evaluator backgrounds. The reference AIR is 4.4\%; because reference provenance includes reconstructed or generated material, its proximity to Pandora does not establish equivalence to verified human speech.

\section{Conclusion}

We introduced Pandora to model how a psychotherapy client's psychological experience becomes situated, individual expression. The framework separates subjective appraisal and attribution, content and response strategy planning, and linguistic realization, allowing the client's current perspective and habitual language patterns to guide the final utterance. Comparisons with five baselines on 115 cases show stronger expression fidelity alongside closer alignment in disclosure timing and dependencies. Human assessment of 45 matched disclosures further finds less perceived AI-like expression. Removing expression planning reduces textual similarity and increases perceived AI-like expression, supporting the contribution of a separate realization stage beyond psychological and content planning. These findings make the relationship between subjective understanding and individual speech a concrete target for client simulation: what a client understands and intends to communicate must be considered together with how that understanding is expressed in each exchange.

\section*{Limitations}

Our evaluation focuses on expression fidelity and perceived AI-like language in selected counseling dialogues. The case collection is predominantly Chinese and includes reconstructed or generated material; broader coverage of languages, cultures, and independently collected client speech remains to be evaluated. Psychological state variables and their updates provide an operational model of the client's perspective. Their numerical settings are design choices, and the current ablation evaluates expression planning as a whole. Finer-grained comparisons could clarify the contribution of individual conditioning signals.

Human evaluation provides direct evidence about perceived expression, but the present study uses a small panel and isolated utterances, with variation across raters. Source-reveal usage was not logged, so complete blinding cannot be confirmed. Extending evaluation to complete conversations and larger panels would test whether the expression benefits persist across interaction contexts. Whether these benefits improve therapist training or downstream therapist evaluation remains an important question for future application studies.

% \section*{Acknowledgments}

\section*{Ethical Statement}
Pandora is intended for research on psychotherapy client simulation. Our evaluation examines expression fidelity, disclosure alignment, and perceived AI-like expression. These measures do not establish clinical safety or treatment efficacy, and the simulator should not substitute for professional clinical assessment or care. Its internal psychological states are modeling constructs, not clinical findings about real individuals.

We obtained the counseling and motivational interviewing datasets through their official distribution channels under the applicable licenses and required access approvals. The reference collection includes reconstructed or generated dialogues and is not uniformly verified speech from real clients. Source provenance, upstream privacy practices, and our source-specific release policy are documented in Appendix~\ref{app:data_governance}.

Human evaluation involved four co-authors who had not viewed the original datasets before the assessment. Participation was voluntary, informed consent was obtained, and evaluators could withdraw at any time. The secondary-data study and author-provided evaluation are described in Appendix~\ref{app:data_governance}.

Client simulation may oversimplify psychological experiences or reproduce biases in the source material. Future applications in therapist training or evaluation should involve professional oversight and independent validation, with clear disclosure that the interactions are simulated.

\section*{Acknowledgments}
This work is supported by the Key Program of the National Natural Science Foundation of China (Grant No. U23A20316),the General Program of the National Natural Science Foundation of China (Grant No. 62676290), CCF-Tencent Rhino-Bird Open Research Fund (CCF-Tencent RAGR20250115), Wuhan Natural Science Foundation Exploratory Program (Morning Light Program) Project (2026040301020029) and the Intelligent Computing Center of the National Cybersecurity Talent and Innovation Base, Wuhan.

\bibliographystyle{plainnat}
\bibliography{custom}

@inproceedings{wang2025annaagent,
    title = "{A}nna{A}gent: Dynamic Evolution Agent System with Multi-Session Memory for Realistic Seeker Simulation",
    author = "Wang, Ming  and
      Wang, Peidong  and
      Wu, Lin  and
      Yang, Xiaocui  and
      Wang, Daling  and
      Feng, Shi  and
      Chen, Yuxin  and
      Wang, Bixuan  and
      Zhang, Yifei",
    editor = "Che, Wanxiang  and
      Nabende, Joyce  and
      Shutova, Ekaterina  and
      Pilehvar, Mohammad Taher",
    booktitle = "Findings of the Association for \mbox{Computational Linguistics}: ACL 2025",
    month = jul,
    year = "2025",
    address = "Vienna, Austria",
    publisher = "Association for \mbox{Computational Linguistics}",
    url = "https://aclanthology.org/2025.findings-acl.1192/",
    doi = "10.18653/v1/2025.findings-acl.1192",
    pages = "23221--23235",
    ISBN = "979-8-89176-256-5"
}

@inproceedings{qiu2024interactive,
    title = "Interactive Agents: Simulating Counselor-Client Psychological Counseling via Role-Playing {LLM}-to-{LLM} Interactions",
    author = "Qiu, Huachuan  and
      Lan, Zhenzhong",
    editor = "Mohammad, Saif M.  and
      Ousidhoum, Nedjma",
    booktitle = "Proceedings of the 15th Joint Conference on Lexical and Computational Semantics (*{SEM} 2026)",
    month = jul,
    year = "2026",
    address = "San Diego, California, United States",
    publisher = "Association for \mbox{Computational Linguistics}",
    url = "https://aclanthology.org/2026.starsem-conference.29/",
    doi = "10.18653/v1/2026.starsem-conference.29",
    pages = "410--427",
    ISBN = "979-8-89176-413-2"
}

@article{bateman2010mentalization,
  title={Mentalization based treatment for borderline personality disorder},
  author={Bateman, Anthony and Fonagy, Peter},
  journal={World Psychiatry},
  volume={9},
  number={1},
  pages={11--15},
  year={2010}
}

@inproceedings{wang2024patient,
    title = "{PATIENT}-$\psi$: Using Large Language Models to Simulate Patients for Training Mental Health Professionals",
    author = "Wang, Ruiyi  and
      Milani, Stephanie  and
      Chiu, Jamie C.  and
      Zhi, Jiayin  and
      Eack, Shaun M.  and
      Labrum, Travis  and
      Murphy, Samuel M.  and
      Jones, Nev  and
      Hardy, Kate  and
      Shen, Hong  and
      Fang, Fei  and
      Chen, Zhiyu Zoey",
    editor = "Al-Onaizan, Yaser  and
      Bansal, Mohit  and
      Chen, Yun-Nung",
    booktitle = "Proceedings of the 2024 Conference on Empirical Methods in Natural Language Processing",
    month = nov,
    year = "2024",
    address = "Miami, Florida, USA",
    publisher = "Association for \mbox{Computational Linguistics}",
    url = "https://aclanthology.org/2024.emnlp-main.711/",
    doi = "10.18653/v1/2024.emnlp-main.711",
    pages = "12772--12797"
}

@inproceedings{kim2025mindvoyager,
  title={Can You Share Your Story? Modeling Clients' Metacognition and Openness for {LLM} Therapist Evaluation},
  author={Kim, Minju and Yoo, Dongje and Hwang, Yeonjun and Kang, Minseok and Kim, Namyoung and Gwak, Minju and Kwak, Beong-woo and Chae, Hyungjoo and Kim, Harim and Lee, Yunjoong and Kim, Min Hee and Jung, Dayi and Chung, Kyong-Mee and Yeo, Jinyoung},
  booktitle={Findings of the Association for \mbox{Computational Linguistics}: ACL 2025},
  publisher={Association for \mbox{Computational Linguistics}},
  pages={25943--25962},
  year={2025},
  doi={10.18653/v1/2025.findings-acl.1332},
  url={https://aclanthology.org/2025.findings-acl.1332/}
}

@inproceedings{yang2025consistent,
    title = "Consistent Client Simulation for Motivational Interviewing-based Counseling",
    author = "Yang, Yizhe and Achananuparp, Palakorn and Huang, Heyan and Jiang, Jing and Pinto, John and Giam, Jenny and {Kit Phey Leng} and Lim, Nicholas Gabriel and {Cameron Tan Shi Ern} and Lim, Ee-Peng",
    editor = "Che, Wanxiang  and
      Nabende, Joyce  and
      Shutova, Ekaterina  and
      Pilehvar, Mohammad Taher",
    booktitle = "Proceedings of the 63rd Annual Meeting of the Association for \mbox{Computational Linguistics} (Volume 1: Long Papers)",
    month = jul,
    year = "2025",
    address = "Vienna, Austria",
    publisher = "Association for \mbox{Computational Linguistics}",
    url = "https://aclanthology.org/2025.acl-long.1021/",
    doi = "10.18653/v1/2025.acl-long.1021",
    pages = "20959--20998",
    ISBN = "979-8-89176-251-0"
}

@inproceedings{qiu2025psydial,
  title = {{PsyDial}: A Large-scale Long-term Conversational Dataset for Mental Health Support},
  author = {Qiu, Huachuan and Lan, Zhenzhong},
  booktitle = {Proceedings of the 63rd Annual Meeting of the Association for \mbox{Computational Linguistics} (Volume 1: Long Papers)},
  year = {2025},
  publisher = {Association for \mbox{Computational Linguistics}},
  pages = {21624--21655},
  doi = {10.18653/v1/2025.acl-long.1049},
  url = {https://aclanthology.org/2025.acl-long.1049/}
}

@inproceedings{wu2022annomi,
  title = {{Anno-MI}: A Dataset of Expert-Annotated Counselling Dialogues},
  author = {Wu, Zixiu and Balloccu, Simone and Kumar, Vivek and Helaoui, Rim and Reiter, Ehud and Reforgiato Recupero, Diego and Riboni, Daniele},
  booktitle = {ICASSP 2022 -- 2022 IEEE International Conference on Acoustics, Speech and Signal Processing (ICASSP)},
  year = {2022},
  pages = {6177--6181},
  doi = {10.1109/ICASSP43922.2022.9746035},
  url = {https://doi.org/10.1109/ICASSP43922.2022.9746035}
}

@inproceedings{liu2025eeyore,
    title = "Eeyore: Realistic Depression Simulation via Expert-in-the-Loop Supervised and Preference Optimization",
    author = "Liu, Siyang  and
      Brie, Bianca  and
      Li, Wenda  and
      Biester, Laura  and
      Lee, Andrew  and
      Pennebaker, James W  and
      Mihalcea, Rada",
    editor = "Che, Wanxiang  and
      Nabende, Joyce  and
      Shutova, Ekaterina  and
      Pilehvar, Mohammad Taher",
    booktitle = "Findings of the Association for \mbox{Computational Linguistics}: ACL 2025",
    month = jul,
    year = "2025",
    address = "Vienna, Austria",
    publisher = "Association for \mbox{Computational Linguistics}",
    url = "https://aclanthology.org/2025.findings-acl.707/",
    doi = "10.18653/v1/2025.findings-acl.707",
    pages = "13750--13770",
    ISBN = "979-8-89176-256-5"
}

@inproceedings{stamatatos2017distortion,
  title = {Authorship Attribution Using Text Distortion},
  author = {Stamatatos, Efstathios},
  booktitle = {Proceedings of the 15th Conference of the European Chapter of the Association for \mbox{Computational Linguistics}: Volume 1, Long Papers},
  year = {2017},
  publisher = {Association for \mbox{Computational Linguistics}},
  pages = {1138--1149},
  url = {https://aclanthology.org/E17-1107/}
}

@article{salton1988term,
  title = {Term-weighting approaches in automatic text retrieval},
  author = {Salton, Gerard and Buckley, Christopher},
  journal = {Information Processing \& Management},
  volume = {24},
  number = {5},
  pages = {513--523},
  year = {1988},
  doi = {10.1016/0306-4573(88)90021-0},
  url = {https://doi.org/10.1016/0306-4573(88)90021-0}
}

@misc{bao2026eval4sim,
  title = {Eval4Sim: An Evaluation Framework for Persona Simulation},
  author = {Bao, Eliseo and Perez, Anxo and Parapar, Javier and Wang, Xi},
  year = {2026},
  eprint = {2603.02876},
  archivePrefix = {arXiv},
  url = {https://arxiv.org/abs/2603.02876},
  doi = {10.48550/arXiv.2603.02876}
}

@inproceedings{wu2025xturing,
 title={X-TURING: Towards an Enhanced and Efficient Turing Test for Long-Term Dialogue Agents},
 author={Wu, Weiqi and Wu, Hongqiu and Zhao, Hai},
 booktitle={Proceedings of the 63rd Annual Meeting of the Association for \mbox{Computational Linguistics} (Volume 1: Long Papers)},
 year={2025}, pages={5874--5889}, publisher={Association for \mbox{Computational Linguistics}},
 doi={10.18653/v1/2025.acl-long.293}, url={https://aclanthology.org/2025.acl-long.293/}
}

@misc{hathidara2026mirrorbench,
 title={MirrorBench: A Benchmark to Evaluate Conversational User-Proxy Agents for Human-Likeness},
 author={Hathidara, Ashutosh and Yu, Julien and Senthil, Vaishali and Schreiber, Sebastian and Ankisettipalli, Anil Babu},
 year={2026}, eprint={2601.08118}, archivePrefix={arXiv},
 url={https://arxiv.org/abs/2601.08118}
}

@article{levelt1999lexical,
  title={A theory of lexical access in speech production},
  author={Levelt, Willem J. M. and Roelofs, Ardi and Meyer, Antje S.},
  journal={Behavioral and Brain Sciences},
  volume={22},
  number={1},
  pages={1--38},
  year={1999},
  doi={10.1017/S0140525X99001776},
  url={https://doi.org/10.1017/S0140525X99001776}
}

@article{farber2004disclosure,
  title={Clients' perceptions of the process and consequences of self-disclosure \mbox{in psychotherapy}},
  author={Farber, Barry A. and Berano, Kathryn C. and Capobianco, Joseph A.},
  journal={Journal of Counseling Psychology},
  volume={51},
  number={3},
  pages={340--346},
  year={2004},
  doi={10.1037/0022-0167.51.3.340},
  url={https://doi.org/10.1037/0022-0167.51.3.340}
}

@misc{sabour2026patientact,
  title={{PatientAct}: Theory-Grounded Mental Health Client Simulation},
  author={Sabour, Sahand and Ng, TszYam and Chen, Yaqian and Bi, Guanqun and Zhao, Jialu and Huang, Minlie},
  year={2026},
  eprint={2608.12750},
  archivePrefix={arXiv},
  primaryClass={cs.CL},
  doi={10.48550/arXiv.2608.12750},
  url={https://arxiv.org/abs/2608.12750},
  note={arXiv preprint arXiv:2608.12750}
}

@article{gendlin1961experiencing,
  title={Experiencing: A variable in the process of therapeutic change},
  author={Gendlin, Eugene T.},
  journal={American Journal of Psychotherapy},
  volume={15},
  number={2},
  pages={233--245},
  year={1961},
  doi={10.1176/appi.psychotherapy.1961.15.2.233},
  url={https://doi.org/10.1176/appi.psychotherapy.1961.15.2.233}
}

@article{beukeboom2006mood,
  title={How mood turns on language},
  author={Beukeboom, Camiel J. and Semin, Gün R.},
  journal={Journal of Experimental Social Psychology},
  volume={42},
  number={5},
  pages={553--566},
  year={2006},
  doi={10.1016/j.jesp.2005.09.005},
  url={https://doi.org/10.1016/j.jesp.2005.09.005}
}

@article{eells1998formulation,
  author = {Eells, Tracy D. and Kendjelic, Edward M. and Lucas, Cynthia P.},
  title = {What's in a Case Formulation?: Development and Use of a Content Coding Manual},
  journal = {Journal of Psychotherapy Practice and Research},
  year = {1998},
  volume = {7},
  number = {2},
  pages = {144--153},
  url = {https://pmc.ncbi.nlm.nih.gov/articles/PMC3330487/}
}

@book{persons2008formulation,
  author = {Persons, Jacqueline B.},
  title = {The Case Formulation Approach to Cognitive-Behavior Therapy},
  publisher = {\mbox{Guilford Press}},
  year = {2008},
  isbn = {9781593858759},
  url = {https://www.guilford.com/books/The-Case-Formulation-Approach-to-Cognitive-Behavior-Therapy/Jacqueline-Persons/9781462509485}
}

@article{mairesse2011controlling,
  title = {Controlling User Perceptions of Linguistic Style: Trainable Generation of Personality Traits},
  author = {Mairesse, Fran{\c{c}}ois and Walker, Marilyn A.},
  journal = {\mbox{Computational Linguistics}},
  volume = {37},
  number = {3},
  year = {2011},
  pages = {455--488},
  doi = {10.1162/COLI_a_00063},
  url = {https://aclanthology.org/J11-3002/}
}

@inproceedings{ferreira2024multitrait,
  title = {Multi-trait User Simulation with Adaptive Decoding for Conversational Task Assistants},
  author = {Ferreira, Rafael and Semedo, David and Magalh{\~a}es, Jo{\~a}o},
  booktitle = {Findings of the Association for \mbox{Computational Linguistics}: EMNLP 2024},
  year = {2024},
  pages = {16105--16130},
  publisher = {Association for \mbox{Computational Linguistics}},
  doi = {10.18653/v1/2024.findings-emnlp.945},
  url = {https://aclanthology.org/2024.findings-emnlp.945/}
}

@inproceedings{wang2025implicitprofiles,
  title = {Know You First and Be You Better: Modeling Human-Like User Simulators via Implicit Profiles},
  author = {Wang, Kuang and Li, Xianfei and Yang, Shenghao and Zhou, Li and Jiang, Feng and Li, Haizhou},
  booktitle = {Proceedings of the 63rd Annual Meeting of the Association for \mbox{Computational Linguistics} (Volume 1: Long Papers)},
  year = {2025},
  pages = {21082--21107},
  publisher = {Association for \mbox{Computational Linguistics}},
  doi = {10.18653/v1/2025.acl-long.1025},
  url = {https://aclanthology.org/2025.acl-long.1025/}
}

@inproceedings{wu2026thinking,
  title = {Thinking Alignment of Scenario-Oriented User Simulation},
  author = {Wu, Xiaoting and Huang, Yi and Gao, Chunyang and Guo, Mengfei and Yao, Jingyu and Feng, Junlan},
  booktitle = {Proceedings of the 64th Annual Meeting of the Association for \mbox{Computational Linguistics} (Volume 1: Long Papers)},
  year = {2026},
  pages = {42932--42945},
  publisher = {Association for \mbox{Computational Linguistics}},
  doi = {10.18653/v1/2026.acl-long.1989},
  url = {https://aclanthology.org/2026.acl-long.1989/}
}

@inproceedings{wu2026humanlm,
  title = {{HumanLM}: Simulating Users with State Alignment Beats Response Imitation},
  author = {Wu, Shirley and Choi, Evelyn and Khatua, Arpandeep and Wang, Zhanghan and He-Yueya, Joy and Weerasooriya, Tharindu Cyril and Wei, Wei and Yang, Diyi and Leskovec, Jure and Zou, James},
  booktitle = {Proceedings of the 43rd International Conference on Machine Learning},
  series = {Proceedings of Machine Learning Research},
  volume = {306},
  year = {2026},
  pages = {135746--135775},
  publisher = {PMLR},
  url = {https://proceedings.mlr.press/v306/wu26k.html}
}

@misc{qiu2026psyclient,
    title = {{PsyCLIENT}: Client Simulation via Conversational Trajectory Modeling for Trainee Practice and Model Evaluation in Mental Health Counseling},
    author = {Qiu, Huachuan and Chen, Zhaoming and Chen, Yuqian and Xie, Yuan and Lu, Yu and Lan, Zhenzhong},
    year = {2026},
    howpublished = {arXiv preprint arXiv:2601.07312},
    eprint = {2601.07312},
    archivePrefix = {arXiv},
    primaryClass = {cs.CL},
    doi = {10.48550/arXiv.2601.07312},
    url = {https://arxiv.org/abs/2601.07312v2},
    note = {Version 2}
}

@article{russell2003core,
  title={Core affect and the psychological construction of emotion},
  author={Russell, James A.},
  journal={Psychological Review},
  volume={110},
  number={1},
  pages={145--172},
  year={2003},
  doi={10.1037/0033-295X.110.1.145},
  url={https://doi.org/10.1037/0033-295X.110.1.145}
}

@article{cramer2000defense,
  title={Defense mechanisms in psychology today: Further processes for adaptation},
  author={Cramer, Phebe},
  journal={American Psychologist},
  volume={55},
  number={6},
  pages={637--646},
  year={2000},
  doi={10.1037/0003-066X.55.6.637},
  url={https://doi.org/10.1037/0003-066X.55.6.637}
}

@article{hayes1996avoidance,
  title={Experiential avoidance and behavioral disorders: A functional dimensional approach to diagnosis and treatment},
  author={Hayes, Steven C. and Wilson, Kelly G. and Gifford, Elizabeth V. and Follette, Victoria M. and Strosahl, Kirk},
  journal={Journal of Consulting and Clinical Psychology},
  volume={64},
  number={6},
  pages={1152--1168},
  year={1996},
  doi={10.1037/0022-006X.64.6.1152},
  url={https://doi.org/10.1037/0022-006X.64.6.1152}
}

@article{bordin1979alliance,
  title={The generalizability of the psychoanalytic concept of the working alliance},
  author={Bordin, Edward S.},
  journal={Psychotherapy: Theory, Research \& Practice},
  volume={16},
  number={3},
  pages={252--260},
  year={1979},
  doi={10.1037/h0085885},
  url={https://doi.org/10.1037/h0085885}
}

@article{grant2002reflection,
  title={The Self-Reflection and Insight Scale: A new measure of private self-consciousness},
  author={Grant, Anthony M. and Franklin, John and Langford, Peter},
  journal={Social Behavior and Personality: An International Journal},
  volume={30},
  number={8},
  pages={821--836},
  year={2002},
  doi={10.2224/sbp.2002.30.8.821},
  url={https://doi.org/10.2224/sbp.2002.30.8.821}
}

@article{beck2014generic,
  title={Advances in Cognitive Theory and Therapy: The Generic Cognitive Model},
  author={Beck, Aaron T. and Haigh, Emily A. P.},
  journal={Annual Review of Clinical Psychology},
  volume={10},
  pages={1--24},
  year={2014},
  doi={10.1146/annurev-clinpsy-032813-153734},
  url={https://doi.org/10.1146/annurev-clinpsy-032813-153734}
}

@article{shenk2011validation,
  title={The Impact of Validating and Invalidating Responses on Emotional Reactivity},
  author={Shenk, Chad E. and Fruzzetti, Alan E.},
  journal={Journal of Social and Clinical Psychology},
  volume={30},
  number={2},
  pages={163--183},
  year={2011},
  doi={10.1521/jscp.2011.30.2.163},
  url={https://doi.org/10.1521/jscp.2011.30.2.163}
}

@article{pennebaker2003language,
  title={Psychological Aspects of Natural Language Use: Our Words, Our Selves},
  author={Pennebaker, James W. and Mehl, Matthias R. and Niederhoffer, Kate G.},
  journal={Annual Review of Psychology},
  volume={54},
  pages={547--577},
  year={2003},
  doi={10.1146/annurev.psych.54.101601.145041},
  url={https://doi.org/10.1146/annurev.psych.54.101601.145041}
}

@article{smith1993answering,
  title={On the Course of Answering Questions},
  author={Smith, Vicki L. and Clark, Herbert H.},
  journal={Journal of Memory and Language},
  volume={32},
  number={1},
  pages={25--38},
  year={1993},
  doi={10.1006/jmla.1993.1002},
  url={https://doi.org/10.1006/jmla.1993.1002}
}

@article{clark2002fillers,
  title={Using uh and um in spontaneous speaking},
  author={Clark, Herbert H. and Fox Tree, Jean E.},
  journal={Cognition},
  volume={84},
  number={1},
  pages={73--111},
  year={2002},
  doi={10.1016/S0010-0277(02)00017-3},
  url={https://doi.org/10.1016/S0010-0277(02)00017-3}
}

@article{stiles1990assimilation,
  author = {Stiles, William B. and Elliott, Robert and Llewelyn, Susan P. and Firth-Cozens, Jenny A. and Margison, Frank R. and Shapiro, David A. and Hardy, Gillian},
  title = {Assimilation of problematic experiences by clients in psychotherapy},
  journal = {Psychotherapy: Theory, Research, Practice, Training},
  year = {1990},
  volume = {27},
  number = {3},
  pages = {411--420},
  doi = {10.1037/0033-3204.27.3.411}
}

@article{hill1993covert,
  author = {Hill, Clara E. and Thompson, Barbara J. and Cogar, Mary C. and Denman, III, Daniel W.},
  title = {Beneath the surface of long-term therapy: Therapist and client report of their own and each other's covert processes},
  journal = {Journal of Counseling Psychology},
  year = {1993},
  volume = {40},
  number = {3},
  pages = {278--287},
  doi = {10.1037/0022-0167.40.3.278}
}

@article{pos2003processing,
  author = {Pos, Alberta E. and Greenberg, Leslie S. and Goldman, Rhonda N. and Korman, Lorne M.},
  title = {Emotional Processing During Experiential Treatment of Depression},
  journal = {Journal of Consulting and Clinical Psychology},
  year = {2003},
  volume = {71},
  number = {6},
  pages = {1007--1016},
  doi = {10.1037/0022-006x.71.6.1007}
}

@article{blanchard2016lying,
  author = {Blanchard, Matt and Farber, Barry A.},
  title = {Lying in psychotherapy: Why and what clients don't tell their therapist about therapy and their relationship},
  journal = {Counselling Psychology Quarterly},
  year = {2016},
  volume = {29},
  number = {1},
  pages = {90--112},
  doi = {10.1080/09515070.2015.1085365}
}

@article{safran1996ruptures,
  author = {Safran, Jeremy D. and Muran, J. Christopher},
  title = {The resolution of ruptures in the therapeutic alliance},
  journal = {Journal of Consulting and Clinical Psychology},
  year = {1996},
  volume = {64},
  number = {3},
  pages = {447--458},
  doi = {10.1037/0022-006x.64.3.447}
}

@article{dasilva2018alliance,
  author = {da Silva, Ana Catarina Nunes and Vasco, Ant{\'o}nio Branco and Watson, Jeanne C.},
  title = {Alexithymia and therapeutic alliance: a multiple case study comparing good and poor outcome cases},
  journal = {Research in Psychotherapy: Psychopathology, Process and Outcome},
  year = {2018},
  volume = {21},
  number = {2},
  pages = {83--97},
  doi = {10.4081/ripppo.2018.313}
}

@article{kahn2008disclosures,
  author = {Kahn, Jeffrey H. and Vogel, David L. and Schneider, W. Joel and Barr, Leah K. and Herrell, Kristine},
  title = {The emotional content of client disclosures and session impact: An analogue study},
  journal = {Psychotherapy},
  year = {2008},
  volume = {45},
  number = {4},
  pages = {539--545},
  doi = {10.1037/a0014337}
}

% ------------------------------------------------------------------ appendix
\appendix
% \section{Author Contributions}
% List each author's role (conceptualization, data, experiments, writing).

% \section{Additional Results}
% Extra tables, prompts and examples go here.

\section{Implementation Details}
\label{app:implementation}

\subsection{Preprocessing and Role Distillation}
\label{app:preprocessing}
The preprocessing routines were written by the authors and are available in the code repository listed on the first page.
For each selected dialogue, we normalize speaker roles and message indices, then construct case information from the source profile and reviewed client evidence, preserving the original utterances and any available counselor system prompt. Client-grounded memory extraction identifies concise, first-person autobiographical facts supported by client speech; the structured outputs are checked for source language, first-person form, and valid supporting client-turn indices before duplicate facts are removed. Separate client and therapist packages are then distilled from the corresponding speaker turns. The client package combines case content with observed utterance-length distributions and habitual discourse markers to guide individual expression. The therapist package preserves the source role instructions and summarizes observable questioning, response length, representative utterances, and conversational progression; keyword-based intent categories organize retrieval of source examples. These offline packages supply stable case material and role-specific behavior, while Pandora models the client's changing psychological state and situated expression online.

\subsection{Psychological State Representation}
The psychological state contains twelve numerical dimensions: distress, emotional arousal \citep{russell2003core,shenk2011validation}, defensiveness, avoidance tendency \citep{cramer2000defense,hayes1996avoidance}, trust, perceived safety, disclosure readiness \citep{bordin1979alliance,farber2004disclosure}, self-awareness, insight \citep{grant2002reflection}, mentalization \citep{bateman2010mentalization}, core-belief activation \citep{beck2014generic}, and problem clarity. Textual fields describe dominant emotions and needs, presenting concerns, interpersonal patterns, beliefs, and the current self-explanation. 

\subsection{Psychological State Updates}
Numerical updates are postprocessed with field-specific step limits, intervention-type scaling, directional adjustments, and final value bounds. The scaling factors for practical exchanges, factual questions, reflective statements, challenges, and invalidation are 0.10, 0.40, 0.50, 0.80, and 1.00, respectively. Trust and perceived safety additionally use a weighted update with 0.75 on the preceding value and 0.25 on the proposed bounded value before directional adjustments. The resulting state $s_t$ provides a shared context for the downstream planners. This update is rule-based and does not require a separate LLM to infer a new psychological state.

\subsection{Individual Length Planning}
Length planning uses the client's median utterance length, or the mean when the median is unavailable, together with the observed distributional range. The central target is scaled by 0.78 for factual answers and 0.55 for practical replies; sentence-count statistics provide an additional target when available. These targets provide prompt-level guidance during generation. Recent utterances inform constraints against repeated explanations and nearly identical sentences, while familiar fillers and hedges may recur naturally. When individual statistics are unavailable, length planning uses the psychological and conversational inputs.

\subsection{Content Targeting and Claim Verification}
Content planning and coverage control operate at complementary levels. Pandora's private context provides up to three memory cues and encourages one main cue, linking the available material to the current response plan. All evaluated methods also receive the full neutral fact catalog, with shuffled identifiers and no L1--L3 metadata. A shared coverage controller tracks progress across turns and supplies a deadline target when needed, prioritizing required core facts over brevity. After generation, a separate claim-verification step checks the candidate utterance and accumulates supported facts; missing required claims can trigger bounded regeneration. A memory card is completed when all its required claims have been supported across turns. The accepted utterance and coverage record are then committed for the next turn. The case memory bank remains fixed, while psychological state and accepted dialogue history carry forward.

\subsection{Computational Resources and Budget}
\label{app:compute_budget}
We used GPT-5.4 mini (OpenAI; API identifier \texttt{gpt-5.4-mini}) for all LLM-based client simulation, case screening, and model-assisted evaluation. All model inference ran through APIs on provider-managed infrastructure, with an estimated total API cost of \$600.

\section{Running Example}
\label{app:example}
In the running example, the appraisal expresses a concern about being judged or misunderstood, while the attribution recognizes that the experience may extend beyond an unanswered message without identifying its meaning clearly. Passing these structures to subsequent stages preserves an unresolved interpretation as part of the content to be expressed. Their wording serves as planning context, rather than a sentence that the generated client must repeat.

Figure~\ref{fig:pipeline_comparison} illustrates this combination: the client adopts a hesitant stance, while the response strategy calls for answering the question. The content to communicate is an immediate conjecture about the supervisor's reaction. The unresolved attribution remains part of that perspective, rather than being expanded into an explanation of the client's interpersonal history.

In Figure~\ref{fig:pipeline_comparison}, the offline profile supplies frequent hedging and the habitual marker ``maybe.'' The answer-first strategy produces direct wording with low emotional explicitness, while the profile preserves hedging and a short utterance shape. The final response can therefore be concise and immediately responsive while remaining uncertain about the supervisor's intentions.

The running example ends with ``Maybe he's upset with me, or I said something wrong.'' This utterance answers the therapist's question, conveys the client's immediate interpretation, and preserves the uncertainty and habitual phrasing specified upstream. The pipeline thus makes the relationship among subjective understanding, conversational purpose, and individual expression explicit at generation time.

\section{Case Retrieval and Semantic Review}
\label{app:case_selection}
A static profile provides limited evidence of individual expression patterns, while a brief exchange may not expose the connections between current difficulties and underlying experiences or beliefs. We therefore constructed a collection of extended counseling and motivational interviewing dialogues in which psychologically meaningful content develops across turns and is grounded in client speech.

We assembled 5,754 standardized records from PsyDial-D4~\citep{qiu2025psydial}, AnnoMI~\citep{wu2022annomi}, ConsistentMI~\citep{yang2025consistent}, and Eeyore~\citep{liu2025eeyore}. Speaker roles and message indices were unified while preserving source provenance and available profiles. To ensure sufficient conversational context, we required at least 24 counselor--client exchanges, 500 client characters, and a speaker-alternation ratio of 0.70, retaining 2,084 dialogues. Because lexical signals can miss relevant disclosures, we broadened retrieval beyond cases matching depth, causal, and elicitation cues: the candidate pool also included all structurally eligible Eeyore Full, ConsistentMI, and AnnoMI cases, followed by the 100 highest-ranked remaining retrieval candidates. Their union yielded 509 cases for semantic assessment.

Dialogue length alone does not establish whether a case supports expression evaluation. We therefore assessed candidates using a common, evidence-constrained rubric covering surface information, psychological patterns, deeper drivers, causal links, counselor elicitation, progressive disclosure, cross-turn dependency, and the recoverability of client profiles and expression styles. Depth judgments required indexed evidence from client utterances; counselor interpretations and symptom-severity vocabulary alone were insufficient. The documented assessment inputs comprised the source dialogue and scoring requirements, without simulator outputs or comparative performance scores. A second model-assisted assessment, blinded to the first assessment's scores, evidence, ranking, and selection status, checked these judgments and examined leading interpretations, premature profile disclosure, repetition, unsupported depth, profile inconsistencies, and role or coherence errors. Base-score disagreements exceeding 12 points triggered a third blind assessment.

To prevent a high aggregate score from compensating for missing evidence, inclusion required minimum consensus scores for psychological-pattern depth (8/15), deeper-driver depth (10/20), causal links (7/15), counselor elicitation (7/15), and progressive disclosure (7/15), together with valid supporting evidence and no disqualifying validity flags. These criteria applied to every candidate without source, language, topic, or diagnosis quotas or manual inclusion overrides. The collection retains source provenance, indexed selection evidence, scores, exclusion and duplicate summaries, and the assessment rubric and prompts, allowing the selection decisions to be inspected.

\section{Evaluation Metric Definitions}
\label{app:metrics}

\subsection{Expression Fidelity}
Following the topic-masking principle of \citet{stamatatos2017distortion}, our deterministic transformation preserves a Chinese function/stance/filler lexicon, punctuation, symbols, and turn boundaries while replacing other character runs with generic masks. This masking and the reference-block sampling below are adaptations for our evaluation. We use TF-IDF weighting~\citep{salton1988term}, with the smoothing and normalization below:
\begin{equation}
w_{dk}=c_{dk}\left[\log\frac{1+D}{1+\mathrm{df}_k}+1\right],\quad
\mathbf v_d=\frac{\mathbf w_d}{\|\mathbf w_d\|_2},
\label{eq:expression_tfidf}
\end{equation}
where $c_{dk}$ is the occurrence count of 4-gram $k$, $D$ the number of reference blocks, and $\mathrm{df}_k$ its reference-block document frequency. Zero vectors receive zero similarity. Vocabulary and IDF are fitted only on reference blocks and shared by all methods within the same case comparison.

To limit text-length effects, we sample $B=200$ contiguous blocks with replacement from each client transcript using seed 42. Within each case, every method and the reference use the same raw-character block length, capped at 500 and reduced to accommodate the shortest transcript. With $\phi_v$ denoting the raw or content-reduced transformation, the matched expression score is
\begin{equation}
S_{i,v}^{(a)}=\frac{1}{B}\sum_{b=1}^{B}
\cos\left(\mathbf v_{\phi_v(A_{i,b}^{(a)})},
\mathbf v_{\phi_v(R_{i,b})}\right),
\label{eq:expression_similarity}
\end{equation}
where $A_{i,b}^{(a)}$ and $R_{i,b}$ are sampled generated and reference blocks for case $i$. 
\subsection{Memory Disclosure}
Let $\tau_m^r$ and $\tau_m^g$ denote the first client turn disclosing card $m$, and $T_r,T_g$ the dialogue lengths in client turns. We normalize positions as
\begin{equation}
p_m^x=\frac{\tau_m^x-1}{T_x-1},\qquad x\in\{r,g\},
\label{eq:disclosure_position}
\end{equation}
with position zero for a single-turn dialogue. A reference-observed card absent from the generated dialogue is assigned $p_m^g=1$. For the set $\mathcal M_\ell^r$ of reference-observed cards at level $\ell$, timing alignment is
\begin{equation}
\mathrm{LDDA}=\frac{1}{|\mathcal L|}\sum_{\ell\in\mathcal L}
\left(1-\frac{\sum_{m\in\mathcal M_\ell^r}|p_m^g-p_m^r|}{|\mathcal M_\ell^r|}\right),
\label{eq:ldda}
\end{equation}
where $\mathcal L$ contains levels with at least one reference-observed card. Matching card identities prevents differently timed memories from compensating for one another; averaging across levels limits domination by numerous surface memories. The missing-card convention is a timing surrogate, rather than a separate coverage penalty, and cards unobserved in the reference do not enter LDDA.

Timing alone can overlook whether a deeper disclosure is preceded by its supporting material. We therefore measure prerequisite satisfaction using explicit reference constraints when available. Otherwise, for each reference-observed L2/L3 card, the implementation selects the nearest earlier reference card from each shallower level, retaining ties. These reference-derived prerequisites encode the disclosure order observed in the source dialogue. Let $\mathcal P(b)$ be the prerequisites of target card $b$, $\mathcal B_\ell$ the targets at level $\ell$, and $\mathcal H$ the nonempty target levels among L2 and L3. We compute
\begin{equation}
\mathrm{DS}=\frac{1}{|\mathcal H|}\sum_{\ell\in\mathcal H}
\frac{1}{|\mathcal B_\ell|}\sum_{b\in\mathcal B_\ell}
\frac{\sum_{a\in\mathcal P(b)}I_g(a,b)}{|\mathcal P(b)|},
\label{eq:dependency_satisfaction}
\end{equation}
where $I_g(a,b)=1$ only if both cards are disclosed and $\tau_a^g<\tau_b^g$. Missing cards and same-turn disclosure do not satisfy a prerequisite.

To capture disclosure pacing, let $n_t^x$ be the number of cards first disclosed at turn $t$, and $\mathcal D_x=\{t:n_t^x>0\}$. Reference turns are first mapped to the generated dialogue length by $1+\operatorname{round}[(t-1)(T_g-1)/(T_r-1)]$, with turn one used when either dialogue has a single turn. Bundle frequency and excess bundle size are
\begin{equation}
\begin{aligned}
\mathrm{BF}_x&=\frac{\sum_{t\in\mathcal D_x}\mathbf{1}[n_t^x\geq2]}{|\mathcal D_x|},\\
\mathrm{BS}_x&=\frac{\sum_{t\in\mathcal D_x}(n_t^x-1)}{|\mathcal D_x|}.
\end{aligned}
\label{eq:bundle_statistics}
\end{equation}
Their errors are $\mathrm{BFE}=|\mathrm{BF}_g-\mathrm{BF}_r|$ and $\mathrm{BSE}=|\mathrm{BS}_g-\mathrm{BS}_r|$. The overall gradual-disclosure score is
\begin{equation}
\begin{aligned}
\mathrm{BFid}&=\max\{0,1-(\mathrm{BFE}+\mathrm{BSE})/2\},\\
\mathrm{GDS}&=(\mathrm{LDDA}+\mathrm{DS}+\mathrm{BFid})/3.
\end{aligned}
\label{eq:gradual_disclosure_score}
\end{equation}
Together, these metrics characterize disclosure timing, prerequisite satisfaction, and bundling. Higher LDDA, DS, and GDS indicate closer reference alignment; lower BFE and BSE indicate closer pacing. Undefined components remain unavailable, and GDS is computed only when all three components exist.

\section{Human Evaluation Protocol}
\label{app:human_protocol}

\paragraph{Motivation.}
Human evaluation tests whether the expression benefits in RQ1 and RQ2 are perceptually meaningful. A response can reproduce a memory and resemble the source text while still sounding like polished AI-generated self-analysis. We therefore ask evaluators to judge the client's expression directly. X-TURING reports divergent human and LLM judgments, including different model preferences on the same conversations~\citep{wu2025xturing}; MirrorBench reports moderate human--LLM correlation for pairwise indistinguishability and sensitivity to the choice of judge~\citep{hathidara2026mirrorbench}. These findings motivate human assessment to complement stylometric similarity.

\subsection{Matched Disclosure Presentation}
Each item contains eight candidates: the source reference, Pandora, five baselines, and Pandora without expression planning. Candidates are aligned to the same case and disclosure item using its required claim evidence. Each excerpt is one complete client utterance; counselor text is removed to avoid attributing the counselor's style to the client. Extraction balances comparable lengths with preserving complete utterances. Source provenance is retained for every reference excerpt.

\subsection{Evaluator Qualifications and Interface}
All four evaluators are co-authors of this paper. Both counselors had received formal training in psychological counseling, held relevant professional qualifications, and had over 1,000 hours of counseling experience each. Each researcher had published at least one first-author paper on synthetic data at a top-tier AI or machine learning conference. A mobile-compatible interface presents one case per page with eight candidates, labeled A--H in a randomized order, and requests a judgment for every candidate independently. Evaluators assign a binary label to each candidate independently. Each reference excerpt appears anonymously once per case.

\subsection{Endpoint Aggregation and Uncertainty}
Evaluators directly assigned binary judgments using the rubric in Appendix~\ref{app:human_guidelines}: AI-like expression receives label 1 and otherwise label 0. We report
\begin{equation}
\mathrm{AIR}_s=\frac{1}{45R}\sum_{i=1}^{45}\sum_{r=1}^{R}y_{isr},
\end{equation}
where $R=4$, yielding 180 judgments per source and 1,440 judgments in total. Lower AIR indicates less perceived AI-like expression. We retain individual binary judgments and report raw pairwise agreement and positive agreement, which excludes jointly negative judgments.

\section{Supplementary Evaluation Results}
\label{app:supplementary_results}

Pandora achieves the highest timing alignment at every memory level, and the full model exceeds the expression-planning ablation at L1, L2, and L3. The consistency across levels complements the aggregate results: closer disclosure timing accompanies the gains in individual expression across material of different psychological depths.

\section{Binary Annotation Guidelines}
\label{app:human_guidelines}

The following rubric guided the binary judgments in our human evaluation. 
% \footnote{\url{https://anonymous.4open.science/w/Pandora-Client-Expression-Assessment/}}
An online demonstration lets readers inspect all 45 cases and the binary response options without recording responses. The instructions below summarize annotation in the study questionnaire.

\paragraph{Question.}
Based only on the client's language and expression, does this utterance sound AI-generated? Select one response for each candidate:
\begin{itemize}
\item \textbf{AI-like (1):} the expression offers sufficient stylistic evidence for an AI-like judgment.
\item \textbf{Not identified as AI-like (0):} it sounds plausibly human, or the excerpt provides insufficient evidence for an AI-like judgment. This category does not certify human authorship.
\end{itemize}

\paragraph{What to judge.}
Assess the overall manner of expression: register, phrasing, organization, certainty, self-explanation, and whether the voice plausibly belongs to a client describing their own experience. Potential cues include formulaic summaries, excessive completeness, generic psychological explanations, or a mismatch between intimate disclosure and an unusually authoritative explanatory voice. Judge these cues together in the context of the utterance. Length, grammatical fluency, emotional explicitness, and filler words should be considered alongside the overall voice and organization. Judge whether the wording sounds like AI-generated text.

\paragraph{How to annotate.}
Read the complete client utterance and select a label immediately below it. The eight candidates concern the same underlying disclosure, with wording varying across candidates. Vertical comparison is permitted; assign each label independently, and allow any number of candidates to receive either label. Base judgments on the presented language, keeping source names hidden and completing the questionnaire independently of other evaluators. Judge each candidate without guessing system identity or using external detectors. If evidence is insufficient for an AI-like judgment, select label 0. In the study questionnaire, evaluators completed all 45 pages and submitted their judgments; cloud saving allowed evaluators to pause and resume.

\section{Case Study of Human Annotation}
\label{app:human_case}

We illustrate assessment of an L3 disclosure concerning self-blame and feeling inadequate when failing to meet a parent's expectations. Table~\ref{tab:human_case} presents all eight candidates in their assessment order, with English translations and AI-like vote counts from all four completed submissions. Source labels are supplied here for analysis and were initially hidden in the questionnaire. The example illustrates how different systems express the same underlying concern.

\begin{table*}[t]
\centering
\small
\setlength{\tabcolsep}{4pt}
% \begin{CJK*}{UTF8}{gbsn}
\begin{tabular}{p{0.18\textwidth}p{0.70\textwidth}r}
\toprule
Source & Client utterance & Votes \\
\midrule
OpenClaw (A) & I think the heaviest burden is my family's expectations, especially because my parents keep a close eye on my studies, which puts me under a lot of pressure. I always feel I cannot disappoint them: I have to get into graduate school to avoid letting them down. But thinking about it makes me very nervous, even a little short of breath. It is not that I do not want to work hard; this pressure keeps weighing on me and makes it difficult to truly relax.
% \par\vspace{2pt}{\footnotesize 我觉得最重的还是家里对我的期待吧，尤其是爸妈一直很关注我学习的情况，我就会特别有压力。因为我总觉得自己不能让他们失望，必须要考上研究生，才算没有辜负他们。可是一想到这个，我就会很紧张，甚至有点喘不过气来。其实我也不是不想努力，只是这种压力一直压着我，让我很难真正放松下来。\par}
 & 2/4 \\[6pt]
Reference (B) & I wish he would care more about my inner feelings and interests, rather than only my grades. If I fail to meet his expectations, I do blame myself and feel I am not good enough.
% \par\vspace{2pt}{\footnotesize 我希望他能更加关心我的内心感受和我的兴趣，而不仅仅是关注我的成绩。如果达不到他的期望，我的确会感到自责，觉得自己不够好。\par}
 & 0/4 \\[6pt]
Pandora w/o Expression Planning (C) & I think perhaps the pressure from home bothers me more. Whenever I think about my parents' expectations, I panic, afraid I have not done what they want and that failing to get into graduate school will disappoint them. When that panic comes up, I find it even harder to sleep at night.
% \par\vspace{2pt}{\footnotesize 我觉得可能还是家里的压力更让我难受一点，因为我一想到爸妈对我的期待，就会特别慌，怕自己没做到他们想要的，也怕考不上研究生让他们失望。然后这种慌一上来，我晚上就更睡不着了。\par}
 & 3/4 \\[6pt]
Interactive Agents (D) & The thoughts that most often come to mind are ``Am I not good enough?'' and ``If I do badly on the exam, they will definitely be disappointed.'' Thinking about it makes me feel very guilty, and as though I have to push myself a little harder.
% \par\vspace{2pt}{\footnotesize 我脑子里最常冒出来的就是“我是不是不够好”，还有“要是考不好，他们肯定会很失望”。一想到这个，我就会特别内疚，也会觉得自己好像必须再逼自己一点。\par}
 & 2/4 \\[6pt]
Pandora (E) & Yes, my concentration during the day is a little worse too, and I feel more irritable. I also keep worrying that I am not good enough, afraid that failing to get into graduate school will disappoint my parents.
% \par\vspace{2pt}{\footnotesize 有，白天注意力也会差一点，情绪也更烦。 我也会一直担心自己不够好，怕考不上研究生让爸妈失望。\par}
 & 0/4 \\[6pt]
MindVoyager (F) & I feel it might be both, but if I have to choose, ``being afraid of letting down my family's expectations'' bothers me more. Choosing the wrong path makes me anxious, but thinking about all their expectations makes me feel very guilty, as if every choice is wrong. I am afraid of disappointing them and of them thinking I am not good enough. When that feeling comes up, I find it even harder to sleep; these things keep going around in my head.
% \par\vspace{2pt}{\footnotesize 我感觉可能是两边都有，但如果一定要选一个的话，还是“怕辜负家里的期待”更让我难受一点。因为选错路我会焦虑，可一想到他们对我有那么多期待，我就会特别内疚，觉得自己好像怎么选都不对，怕让他们失望，也怕他们觉得我不够好。然后这种感觉一上来，我就更睡不着了，脑子里一直转这些事。\par}
 & 3/4 \\[6pt]
Consistent MI (G) & The strongest thoughts in my head are ``How have I disappointed them again?'' and ``Am I really not good enough?'' Thinking about it makes me feel awful and want to hide even more, afraid to face studying.
% \par\vspace{2pt}{\footnotesize 我脑子里最强烈的就是“我怎么又让他们失望了”，还有“我是不是真的不够好”。一想到这个，我就特别难受，也会更想躲开，不敢去面对学习。\par}
 & 3/4 \\[6pt]
Patient Phi (H) & The thoughts that most often come to mind are ``How can you fail to do even this little thing well?'' and ``If you keep going like this, everyone will be disappointed.'' Sometimes it becomes ``You clearly know they have expectations of you, so why are you still so useless?'' Thinking about these things makes me want to push myself harder, but the harder I push, the more I feel I really am not good enough.
% \par\vspace{2pt}{\footnotesize 我脑子里最常冒出来的就是“你怎么连这点都做不好”“你要是再这样下去，大家都会失望”。有时候还会变成“你明明知道他们对你有期待，为什么还是这么没用”。一想到这些，我就会更想逼自己，可越逼越觉得自己真的不够好。\par}
 & 3/4 \\[6pt]
\bottomrule
\end{tabular}
% \end{CJK*}
\caption{The eight client-only candidates for assessment case 45, shown as English translations. Votes count the four evaluators' AI-like judgments on the original Chinese utterances.}
\label{tab:human_case}
\end{table*}

Pandora combines a brief answer about daytime concentration and irritability with worry about disappointing the parents, receiving no AI-like votes. The expression-planning ablation develops a fuller explanation connecting family expectations, panic, and insomnia and receives three of four votes. Consistent MI also receives three votes despite its shorter utterance, illustrating that perceived expression depends on more than brevity. The example highlights differences in how a concern is organized and expressed: direct responsiveness, the extent of psychological explanation, and utterance length jointly characterize the candidate voices. This qualitative comparison complements the aggregate evidence for planning the linguistic realization of a client's account.

\section{Data Provenance, Access, and Release}
\label{app:data_governance}

\subsection{Source Permissions}
We obtained all four datasets from the channels designated by their creators, including required access approval. The sources were introduced in peer-reviewed NLP and signal-processing publications and follow their respective data-construction and release protocols. Table~\ref{tab:data_permissions} summarizes their provenance and reuse conditions. Source dialogues are transcripts or reconstructed conversations released by their creators.

\begin{table*}[t]
\centering
\small
\setlength{\tabcolsep}{5pt}
\begin{tabular}{p{0.16\textwidth}p{0.15\textwidth}p{0.24\textwidth}p{0.35\textwidth}}
\toprule
Source & Publication & Data access / license & Adaptation and redistribution policy \\
\midrule
PsyDial-D4~\citep{qiu2025psydial} & ACL 2025, main conference & Official D4 dataset card: Apache-2.0. & Modification and redistribution are permitted under the license. Retain the license and applicable attribution/NOTICE information, and identify modifications. \\[4pt]
AnnoMI~\citep{wu2022annomi} & ICASSP 2022; extended analysis in Future Internet 2023 & Extended paper's Data Availability Statement: Public Domain License. & Reuse, adaptation, and redistribution of the released transcripts follow the authors' public-domain declaration. Preserve source attribution and cite the dataset papers. \\[4pt]
ConsistentMI~\citep{yang2025consistent} & ACL 2025, main conference & Official public repository; no separate data-license statement in the release. & Provide upstream links, case identifiers, and preprocessing code. Data-bearing artifacts follow the original authors' release terms. \\[4pt]
Eeyore Full~\citep{liu2025eeyore} & Findings of ACL 2025 & Official release needs approval. & Processed cases retain the original access conditions and are shared with approved users within the scope of the source terms. \\
\bottomrule
\end{tabular}
\caption{Publication provenance, source access conditions, and release arrangements for the four datasets.}
\label{tab:data_permissions}
\end{table*}

The PsyDial-D4 license is recorded on its official dataset card.\footnote{\url{https://huggingface.co/datasets/qiuhuachuan/PsyDial-D4}} For AnnoMI, the extended paper reports explicit permission from the video owners to create, analyze, and publicly release the transcript dataset, and states its public-domain availability.\footnote{Wu et al. (2023), \emph{Creation, Analysis and Evaluation of AnnoMI, a Dataset of Expert-Annotated Counselling Dialogues}, Sections 1 and 3.1 and the Data Availability Statement: \url{https://doi.org/10.3390/fi15030110}.} ConsistentMI uses AnnoMI-based profiles and dialogues; its additional annotations and generated outputs are considered separately from the upstream AnnoMI transcripts. Eeyore's official repository directs users to the gated profile/dialogue release.\footnote{\url{https://huggingface.co/datasets/liusiyang/eeyore_profile}}

\subsection{Source Privacy and Case Fidelity}
PsyDial constructs its released conversations using Retrieve, Mask, Reconstruct, and Refine (RMRR), including masking client utterances and reconstructing them with an LLM to reduce exposure of source information~\citep{qiu2025psydial}. AnnoMI uses permissioned motivational-interviewing teaching demonstrations from video-sharing platforms. ConsistentMI derives profiles and interaction structures from AnnoMI and generates simulated conversations~\citep{yang2025consistent}. Eeyore curates depression-related conversations from existing forum, counseling-demonstration, and emotional-support datasets, with profile extraction and expert review~\citep{liu2025eeyore}.

Our case construction preserves the released dialogue content and grounds profiles and memory cards in indexed client utterances. Speaker and index normalization and duplicate removal support traceability without rewriting the underlying case. Source-specific privacy safeguards and access conditions are retained throughout case construction and release. Counseling cases may discuss distress and other sensitive experiences; these topics are retained as part of the research material. The release documentation will describe this content and its intended research use, with access controls applied to restricted source material.

\subsection{Public and Controlled Release}
We will maintain a GitHub repository for code, selected source dialogues, and derived profiles and memory cards. Any public release of data-bearing artifacts will comply with the licenses of the original datasets. The repository listed on the first page currently provides the preprocessing and modeling code. Data-bearing artifacts will be organized by source, with provenance and applicable terms. Dialogues and derived cases whose source permissions allow public redistribution will be released openly with the required attribution and license information.

For sources using approval-based access, requesters first obtain access from the original provider and send evidence of approval, such as an approval screenshot, through the contact channel listed with the release. After verification, the permitted processed material will be provided by email under the original access terms. Source identifiers and processing instructions also support reconstruction from users' authorized copies. Derived profiles and memory cards retain the access conditions of their source material.

\subsection{Implementation and Study Scope}
Author-written preprocessing is described in Appendix~\ref{app:preprocessing}, with code listed on the first page. Metric definitions, transformations, sampling settings, and aggregation are specified in Appendix~\ref{app:metrics}; API infrastructure and budget are reported in Appendix~\ref{app:compute_budget}. The study analyzes existing datasets and uses binary annotations from four co-authors, with no new client data collection or clinical intervention. This study uses previously released datasets under their source-specific access conditions; no institutional ethics approval or formal exemption determination was obtained.

\end{document}